\documentclass[11pt]{article}
\usepackage{graphicx}
\usepackage{indentfirst,csquotes}

\usepackage{amssymb,amsthm,amsmath}
\usepackage{xcolor,paralist,hyperref,titlesec,fancyhdr,etoolbox}

\titleformat{\section}
{\normalfont\huge\bfseries\centering}
{\thesection}{10pt}{}

\hypersetup{ colorlinks=true, linkcolor=black, filecolor=black, urlcolor=black }

\usepackage{lipsum}
\usepackage{orcidlink}
\usepackage{algorithm}
\usepackage{algpseudocode}
\usepackage{tikz}
\usepackage{fontenc}
\usepackage{attrib}
\usepackage[T1]{fontenc}
\usepackage[utf8]{inputenc}
\usepackage{mathpazo}

\usepackage{calc}
\usepackage{indentfirst}
\usepackage{setspace}
\usepackage{geometry}
\usepackage{changepage}

\usepackage{fancyhdr}
\usepackage{lastpage}

\usepackage{graphicx}
\usepackage{epstopdf}
\usepackage{tikz}

\usepackage{float}
\usepackage{newfloat}
\usepackage{caption}

\usepackage{amsmath}
\usepackage{amssymb}
\usepackage{amsthm}
\usepackage{upgreek}

\usepackage{booktabs}
\usepackage{multirow}
\usepackage{array}
\usepackage{tabularx}
\usepackage{pbox}
\usepackage{colortbl}

\usepackage{enumitem}

\usepackage{ifthen}
\usepackage{etoolbox}
\usepackage{tabto}
\usepackage{totcount}
\usepackage{attrib}
\usepackage{seqsplit}

\usepackage{xcolor}
\usepackage{soul}

\usepackage{marginnote}
\usepackage{marginfix}

\usepackage{tocloft}

\usepackage{enotez}

\usepackage{lineno}

\usepackage[numbers]{natbib}
\usepackage{url}
\usepackage{hyperref}
\usepackage{cleveref}

\usepackage{scrextend}

\usepackage{draftwatermark}

\begin{document}
\title{Bayesian Model Calibration with Integrated Discrepancy: Addressing Inexact Dislocation Dynamics Models} %%%%%%%%%%%%
\author{
Liam Myhill$^{1*}$,
Enrique Martinez Saez$^{1,2}$,
Sez Russcher$^{1}$\\[6pt]
\small $^{1}$ Department of Mechanical Engineering, Clemson University\\
\small $^{2}$ Department of Materials Science and Engineering, Clemson University\\
\small $^{*}$ \texttt{lmyhill@clemson.edu}
}
\date{\today}
% \address{$^{1}$ \quad Department of Mechanical Engineering, Clemson University \\
% $^{2}$ \quad Department of Materials Science and  Engineering, Clemson University \\ $^{*}$ \quad lmyhill@clemson.edu}
% % \email{lmyhill@clemson.edu}
\maketitle

\let\thefootnote\relax

\begin{abstract}
In this work, a novel approach to Bayesian model calibration routines is developed which reinterprets the traditional definition of model discrepancy as defined by Kennedy and O'Hagan (KOH) \cite{Kennedy2001}. The novelty lies in the integration of $\delta_\theta(x_i)$ GPs within the simulator, which is approximated as a GP surrogate model to ensure computational tractability. This approach assumes that the utilized simulator sufficiently predicts observed trends when calibrated with respect to the application domain, and that all model-form errors can be attributed to uncertainty in the input parameters. In contrast, the KOH method assumes discrepancy to be inherently decoupled from the simulator, acting as a 'catch-all' for various sources of model error. The new method is applied to Molecular Dynamics observations of the critical stress to drive dislocation dipoles, and equivalent predictions using a Discrete Dislocation Dynamics simulator whose coarse-grained physical interpretation of the underlying physical mechanisms requires calibration against MD observations. We present an overview of similar state-aware calibration routines; differentiate the provided approach through redefining the commonly used discrepancy Gaussian process and benchmark against KOH. A philosophical argument as to when application of the proposed method is appropriate is provided, and future directions for expanding upon this methodology are proposed. 
\end{abstract} %%%%%%%%%

\bigskip

\section{Introduction}

% \textbf{\textcolor{red}{Include literature review}}
Bayesian inference  is a widely used approach  for calibrating computer simulations against experimental observations and quantifying uncertainties in model predictions \cite{flynn_bayesian_2019,plumlee_bayesian_2017}. The foundational work of Kennedy and O'Hagan (KOH) \cite{Kennedy2001} provided a framework for the systematic evaluation of computer simulations via Gaussian Process (GP) surrogates and since then has been applied in numerous fields \cite{lei_considering_2020,ling_challenging_nodate,higdon_computer_2008}. The KOH framework entails two decoupled GP models. The first, typically referred to as $\eta(\cdot,\cdot)$, is a computational model that can be replaced with a GP emulator should the computational cost of the simulations make active data collection intractable. The second, $\delta(\cdot)$, describes the discrepancy between the computational model and the observed value from some "ground-truth" experiment that cannot be explained by parameter calibration alone. Over the past two decades, various formulations of this framework have been developed. A recent review paper by Sung and Tuo \cite{sung_review_2024} lists many of the fields where the KOH approach has been successfully utilized along with specialized modifications and alternative methods for various types of applications. 

 The KOH framework, however, is notoriously prone to confounding, where multiple $\theta$ and $\delta(\cdot)$ combinations explain the observed data. In later work, O'Hagan \cite{brynjarsdottir_learning_2014} emphasized the importance of model discrepancy, especially when model emulators are asked to extrapolate beyond the bounds of their training data. It was their conviction that "in order to obtain realistic learning about model parameters, or to extrapolate outside the range of the observations, it is important [...] to incorporate model discrepancy [and] model the prior information about it. \cite{brynjarsdottir_learning_2014}". Maupin \textit{et al.} differentiates discrepancy in spatial and temporal regimes, and highlights the role of differing experimental settings on the quality of observation data. \cite{maupin_model_2020}. Lei \textit{et al.} specifically highlights the role of model inadequacy when addressing uncertainty quantification and model-observation discrepancies, specifically as it applies to the field of cardiology \cite{lei_considering_2020}, where model selection also plays a crucial role in predicting patient outcomes. The sources of model/observation discrepancy are numerous and have complex interdependence \cite{arendt_quantification_2012}. The generally accepted additive treatment of discrepancy is the main cause for the documented identifiability problem in KOH-based Bayesian calibration methods \cite{ling_challenging_nodate} where generating sufficiently large datasets is financially or computationally infeasible. In the limit that the number of observation data and simulation data approach infinity, there is no identifiability issue; however, such limits are unrealistic when dealing with most computationally expensive models. 
 
 % HERE ADD MORE ON WHAT OTHERS HAVE DONE REGARDING DISCREPANCY . 
 
 In this paper, the importance of the discrepancy is explored for problems where the model form is known to be structurally acceptable, but its parameters are known or expected to drift within or through the application domain. With such application in mind, we propose an approach where the model discrepancy term is folded into the simulator and used to correct for within a double-layered GP. This proposed approach is needed for the application in hand where it is assumed \textit{a-priori} that the model physics are adequate to capture observed trends. The objective of the GP emulator is to account for and learn the significant model-form error that originate from the omitted dependency of model parameters across the application domain between atomistic and continuum simulations of line defects in crystalline materials.

\subsection{State of the art calibration}
 Model calibration relative to benchmark observation data is of critical importance, particularly for simulations of physical phenomena where experimental observations are cumbersome to produce. State-of-the-art Bayesian methods involve representing an observed phenomenon, $y(x)$ as a summation of Gaussian Processes (GPs) $\eta(\cdot,\cdot)$ and $\delta_{\eta}(\cdot)$. The KOH formalism is presented in Equation \ref{eq:SoAcal}:
 
 \begin{equation}
    y(x_i)=\eta(x_i,\theta_i)+\delta_{\eta}(x_i)+e_i
    \label{eq:SoAcal}
\end{equation}
 
 The computer model, or simulation $\eta(x,\theta)$, can be considered as a Gaussian Process with nonzero mean function and a covariance provided in Equation \ref{eq:etaCov} \cite{higdon_computer_2008}. The input vector $x_i$ refers to physical quantities which are specific to the $i^{th}$ observed phenomenon, $\theta_i$ are a vector of calibration parameters with variable uncertainty, and $e_i$ are observation errors which are inherent to any experiment. GPs are typically flexible enough to reflect input data even with a zero mean function as the 'kernel' or model covariance encodes the simulator behavior:

\begin{equation}
Cov((x,\theta),(x',\theta'))=\frac{1}{\lambda_\eta}R((x,\theta),(x',\theta');\rho_\eta)
    \label{eq:etaCov}
\end{equation}
where $\{x,\theta\}$ are the sampled physical inputs and calibration parameters utilized in the simulator, and $\{\lambda_\eta,\rho_\eta\}$ denote the marginal precision and correlation lengths of the $\eta$ GP, respectively. The function $R$ is the autocorrelation between input samples to the $\eta$ model, or the various data-points used to train the GP. 

The second GP defines the discrepancy between the observed phenomenon and the model ($\eta$) predictions $\delta_{\eta}(x)$ and is assumed to have zero mean, and a covariance function given by Equation \ref{eq:deltaCov} \cite{higdon_computer_2008}.

\begin{equation}
    Cov(x,x')=\frac{1}{\lambda_\delta}R((x,x');\rho_\delta)
    \label{eq:deltaCov}
\end{equation}
where $\{\lambda_\delta,\lambda_\eta\}$ denote the marginal precision and correlation lengths of the discrepancy $\delta$ GP, respectively. 
Summing the $\eta$ and $\delta_{\eta}$ GPs assumes that all of the model-form error can be decoupled from the simulator predictions, indicating that model deficiency is what necessitates correction of the predicted values.

For most applications where models are in close agreement with observed phenomenon, the Bayesian calibration approach of Kennedy and O'Hagan (KOH) \cite{bayarri_computer_2007} is suitable. Using such a formalism, one is able to account for model-form biases and construct GP emulators which provide metrics of uncertainty quantification that can be used to benchmark the quality of a given model prediction. Such benchmarks are important for evaluating the validity of a model, especially when modeling mechanisms that are critically important for application-based engineering. 

Pitfalls of the KOH formalism include the use of cryptic hyperparameters that lack interpretability and wariness in the application of trained $\delta_\eta(x)$ GPs which can over/underestimate the decoupled model-form error, leading to a lack of trust in critical decision making. In an attempt to address these issues, we propose a modification to the conventional methods that account for model discrepancy as a correction to the input parameters. Equation \ref{eq:intDeltaCal} is an adaptation of the previously formulated state-aware Bayesian calibration techniques, which we shall refer to as the 'integrated delta' formalism:

\begin{equation}
    y(x_i)=\eta(x_i,\theta_i+\delta_{\theta}(x_i))+e_i
    \label{eq:intDeltaCal}
\end{equation}
where $\delta_{\theta}$ are zero-mean GPs acting directly on the input parameters to the $\eta$ model. There exists one $\delta_{\theta}$ GP per calibration parameter, and each varies as a function of the application domain $x_i$. 

We have termed the formulation provided in Equation \ref{eq:intDeltaCal} the 'integrated delta' technique due to the embedding of the $\delta_{\theta}(x_i)$ within the $\eta$ GP. The guiding principle behind this technique is that the constitutive relations governing the simulation behavior are correct, and any form of discrepancy can be attributed to differential values of the input parameters across the application domain. It is hypothesized that this implementation may/will
\begin{itemize}
    \item Allow for extrapolative insight for suitable input parameter values
    \item Increase the interpretability of the model discrepancy
    \item Reduce overfitting of observation data in select cases
    \item Improve the identification of critical calibration parameters $\theta_i$
\end{itemize}

With regards to the extrapolative capabilities of the model given in Equation \ref{eq:intDeltaCal}, it is apparent that its predictive capabilities should improve with respect to Equation $\ref{eq:SoAcal}$. Using traditional methods, when outside the physical application $\{x_i\}$ domain, the summation of $\eta$ and $\delta_\eta$ often result in poor extrapolation \cite{kong_towards_2025}. Typically with KOH, exploration outside of the prior $\{\theta_i\}$ domain is generally forbidden in the Markov-chain Monte Carlo (MCMC) sampling used to update posterior distributions for $\theta_i$. Using the integrated $\delta$ approach, the summation of a fixed $\theta_i$ with $\delta_{\theta}(x_i)$ consistently requires $\eta$ to make predictions based on input parameters outside the prior distributions of $\theta_i$. If the optimal values for $\theta_i$ drift significantly across the application domain, this behavior is ideal and cannot be replicated using a traditional KOH approach. 
% If one carefully preforms the feature engineering and understands the implications of each $\theta_i$, the accuracy of the predictions outside the domain should degrade physically instead of arbitrarily, which is a marked improvement over traditional methods. 
% In the conventional model, $\delta(x)$ is not well-defined outside the application domain, as there exists no observation data in the extrapolated regions.  In the updated context, predictions are governed by the synthetic hyperparameter $\theta+\delta(x)$ such that even if $\delta(x)$ is poorly fitting the observed trends, the emulator prediction ought to converge to the simulator model predictions. 

The modifications of Equation \ref{eq:intDeltaCal} force the observation/simulation discrepancy to interact with the model emulator directly instead of as a correction factor, which drastically improves the interpretability of the model formulation. Calibration parameters remain interpretable as baseline physical parameters while $\delta_\theta(x_i)$ become context dependent parameter variation. The discrepancy influences emulator predictions in a way that reflects the sensitivity of the model, which should tighten posterior distributions for $\theta$ and improve Markov-chain Monte Carlo (MCMC) trajectories. In this formulation, the discrepancy exists in a parameter space that must respect the physics injected into simulator models. When dealing with simulation results which are motivated by physics with the aim of generating constitutive models, more physically motivated discrepancy is attractive. 

Despite the listed model benefits, it is important to note the trade-offs of the proposed modification. Sampling the emulator to make predictions becomes more nonlinear than conventional methods, resulting in an increased computational expense. The identifiability issue, while addressed, is still highly dependent on the model sensitivity. A physical intuition is still required when selecting critical calibration parameters. There is also an elevated importance of the prior hyperparameters for $\delta_\theta(x)$ due to the existence of multiple $\delta_\theta$ GPs, each requiring hyperparameter tuning during the initial MCMC steps; however, implementing adaptive step sizes for individual $\delta_\theta$ mitigate this issue with little added computational cost. 

%%%%%%%%%%%%%%%%%%%%%%%%%%%%%%%%%%%%%%%%%%%%%%%

\subsection{Dislocations and crystal plasticity}

One field that has recently explored verification, validation and uncertainty quantification (VVUQ) and GPs more vigorously is the integrated computational materials engineering (ICME) community. One exciting development which takes advantage of an active learning \cite{williams_batch_2011} approach is the calibration of crystal plasticity flow rules from discrete dislocation dynamics (DDD) simulations. Julian \cite{julian_active_nodate} considers the flow rule determining a crystalline metal's plastic behavior as a GP model. Coupling the GP model with a Bayesian optimization routine to enable active learning allows for a systematic reduction in error with subsequent training cycles. Other work by Akhondzadeh \textit{et al.}\cite{akhondzadeh_dislocation_2020} demonstrates 
how flow rules can be fitted from large DDD datasets; however, their approach makes no attempt at constructing GPs for prior-posterior relationships. This paper outlines how state of the art Bayesian model calibration practices can be applied to mesoscale dislocation dynamics simulations to improve the model agreement relative to atomistic predictions. 

Plastic deformation of crystalline materials is caused by the dynamic behavior of line defects, or dislocations \cite{cai_non-singular_2006,cottrell_thermally_2002,mann_elastic_1949,po_phenomenological_2016}. Dislocations control deformation mechanisms in ductile metals through their nucleation, glide, climb, and interactions with boundaries or other dislocations. Material strengthening, or hardening, occurs when networks of dislocations arrest the motion of other dislocations through a crystal grain, until a critical stress to continue material deformation has been attained. Dislocations have been visualized experimentally via electron microscopy techniques, confirming plasticity theories observed since the early 1930's \cite{taylor_mechanism_1934}. Today, dislocations are studied via coupled experimentation, theory and computer simulation.

One can study dislocations from an atomic perspective using electronic structure methods such as density functional theory (DFT), or molecular dynamics (MD)with empirical interatomic potentials; however, to study dislocation ensembles at larger scales, mesoscale dislocation dynamics models, such as DDD or phase field dislocation dynamics (PFDD) \cite{peng_non-orthogonal_2021} are utilized. DDD is a coarse-grained model which does not have the same physical fidelity as atomistic models; however, it is much less computationally expensive to conduct DDD studies than DFT or MD. To simulate dislocation behavior using DDD, one must first extract important material properties from the atomistic scale such as elastic constants $\{\mu,\nu\}$ and dislocation mobility coefficients $\{B_0^e,B_0^s,B_1^e,B_1^s\}$ \cite{po_phenomenological_2016}. To avoid singularities in the computed stress field about dislocations, a non-singular theory is utilized \cite{cai_non-singular_2006,po_non-singular_2018}. This necessitates further coarse-graining of physical mechanisms and the need to fit an effective core size parameter $l^{C}$. Line tension model parameter $\alpha^{LT}$ are also necessary to maintain physical dislocation shapes and mitigate instabilities arising from the ultraviolet catastrophe \cite{boleininger_ultraviolet_2020}. Each of these variables is an input into the DDD simulation, and is subject to a degree of parameter uncertainty. 

For the context of this work, the main characteristic of dislocations that shall be studied is the critical resolved shear stress $\tau^{CRSS}$ which, when resolved on the dislocation glide plane, results in continuous dislocation glide. A dataset of MD simulations is generated which analyzes $\tau^{CRSS}$ as a function of the separation distance between dislocation glide planes $h^d$. DDD simulations are then instantiated to explore the application domain $\{h^d\}$, and the variables which influence $\tau^{CRSS}$. This subset is identified as $\{l^C,\mu,\nu\}$, as the critical stress is independent of dislocation mobility and there exists no curvature to the dislocation lines in the computation of $\tau^{CRSS}$ for the chosen microstructure. 

The current work is presented as follows: Section \ref{sec:compMethods} outlines developments to conventional calibration routines and how they apply to the defined problem. Subsections \ref{subsec:MDobservations}-\ref{subsec:DDDsampling} define how observation and simulation data is collected. Subsection \ref{subsec:intDelta} introduces the computational methods to deploy the integrated $\delta$ formalism. Section \ref{sec:results} demonstrates the predictive capabilities of the new methodology against the \textit{Gaussian Process Models for Simulation Analysis} (GPM-SA) \cite{gattiker_gaussian_2015} implementation of the KOH formalism for the established DDD dataset. Section \ref{sec:discussion} discusses the ramifications of the new methodology, and proposes several directions for future work. 

% For the context of this work, two important characteristics of dislocations shall be considered. The critical stress to drive dislocation glide, $\tau^{CRSS}$, and the thermal activation rate $\Gamma^{TA}$, shall both be systematically evaluated. Molecular dynamics, being a more physically motivated model, shall be considered as ground-truth observation data, while the DDD model will be calibrated such that the dynamic behavior of DDD dislocations best match the MD observations. 

%%%%%%%%%%%%%%%%%%%%%%%%%%%%%%%%%%%%%%%%%%
\section{Computational methods and model formulation}
\label{sec:compMethods}

The means by which the computational studies are conducted are described in this section. Specific processes for data generation are outlined, along with general descriptions of the utilized models. Section \ref{subsec:intDelta} outlines novel developments to the KOH \cite{brynjarsdottir_learning_2014} formalism, which are similar in many respects to previous "state-aware" approaches \cite{atamturktur_state-aware_2015,plumlee_computer_2019} nested or "deep" GPs \cite{damianou_deep_2013}. 

% \subsection{Motivation and thermal effects}
\subsection{Motivation and critical stress}
\label{subsec:motivation}

The multi-scale modeling of material behavior involves a number of simulation tools capable of spanning several time and length scales. Information is extracted from the smallest scale models of discrete atoms, whose methods suffer from length and timescale limitations, and propagated to larger scale continuum models. In the passing of information between simulation frameworks, model calibration with accompanied uncertainty quantification is critical for benchmarking the predictive capabilities at larger scales. The primary goal of the DDD simulator is to formulate plastic flow-rules, whereby large ensembles of interacting dislocations provide insight on loading configurations and rates to permanently deform engineering metals. 

% One recent breakthrough in the multi-scale modeling objective was the coupling of the viscoplastic self-consistent (VPSC) \cite{zecevic_viscoplastic_2021} framework to ABAQUS finite element software \cite{Abaqus2014}. The VPSC model is effective at extrapolating single-crystal hardening models into polycrystalline hardening models; however, thermal effects are not currently accounted for within VPSC. To account for thermal effects moving forward, MoDELib \cite{po_phenomenological_2016} DDD simulator is being utilized to create datasets of critical stress values as a function of the applied temperature, which are to be exposed to the VPSC and propagated to finite element analysis (FEA) simulations. 

\begin{figure}
    \centering
    \includegraphics[width=0.865\linewidth]{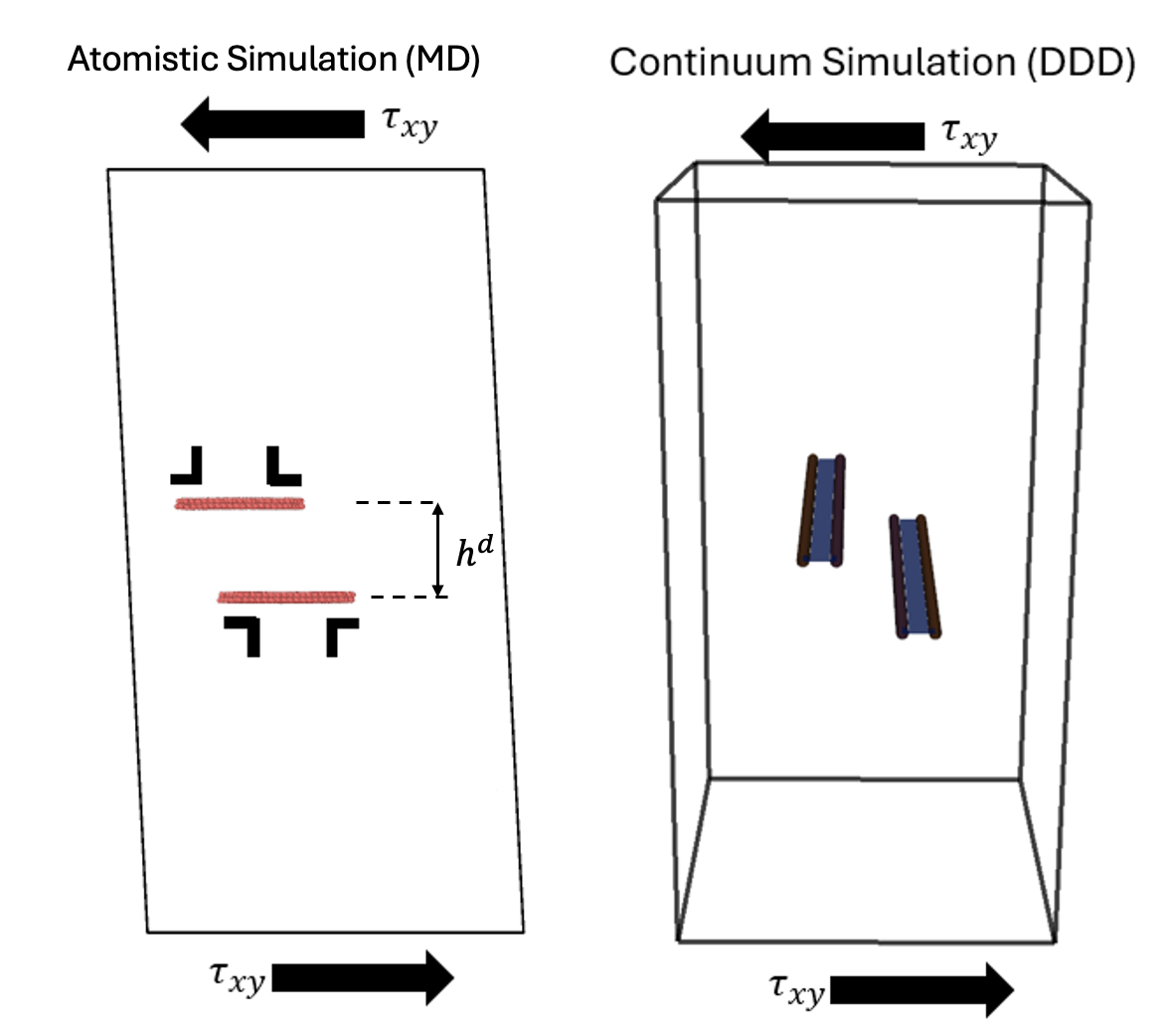}
    \caption{MD (left) and DDD (right) equivalent simulation cells. Dislocation dipoles dissociate into Shockley partials in FCC creating an intrinsic stacking fault colored by the atoms in MD and the shaded field in DDD. The dipole height is the number of glide planes separating dislocation defects.}
    \label{fig:MD-DDDschematic}
\end{figure}

To accurately capture the critical stress to activate large ensembles of interacting dislocations, it is important to first ensure the critical stress necessary to drive a single set of interacting dislocations is consistent between MD and DDD. Dislocation-dislocation interaction is one of the primary mechanisms for material hardening, along with junction formation; however, we do not consider the role of junction formation in this work. The chosen microstructure and loading orientations do not lend themselves to junction formation, due to inability to cross slip for FCC edge Shockley partials whose separation is maintained by a uniform intrinsic staking fault energy. To this effect, a simple dislocation dipole is generated in a face-centered cubic (FCC) Cu MD simulation cell. The simulation cell is replicated in a continuum DDD model, with equivalent dimensions describing the simulation cells and dislocation microstructure. Figure \ref{fig:MD-DDDschematic} depicts the MD simulation cell, and the continuum DDD simulation cell.

% \begin{figure}
%     \centering
%     \includegraphics[width=0.85\linewidth]{Figures/bisection_crss.png}
%     \caption{The bisection search algorithm utilized to determine $\tau_{DDD}^{CRSS}$. A resolution $\varepsilon=5$ MPa was observed.}
%     \label{fig:bisectionSearch}
% \end{figure}

Evaluating the degree of certainty in the predictive capabilities of the chosen DDD simulator, with the goal of verifying the information being passed into larger scales, Bayesian calibration is performed on the DDD model, considering MD simulation results as ground truth observations. Any uncertainty, residual variability, or error in the MD simulations caused by shortcomings in the utilized interatomic potential will be accounted for in the observation error term $e_i$ from Equations \ref{eq:SoAcal} and \ref{eq:intDeltaCal}. The DDD simulation data, $\eta(x,\theta)$, shall be utilized to train a set of embedded GPs $\eta(x,\theta+\delta(x))$ where both $\eta$ and $\delta$ are GPs.

\begin{figure}
    \centering
    \includegraphics[width=1.0\linewidth]{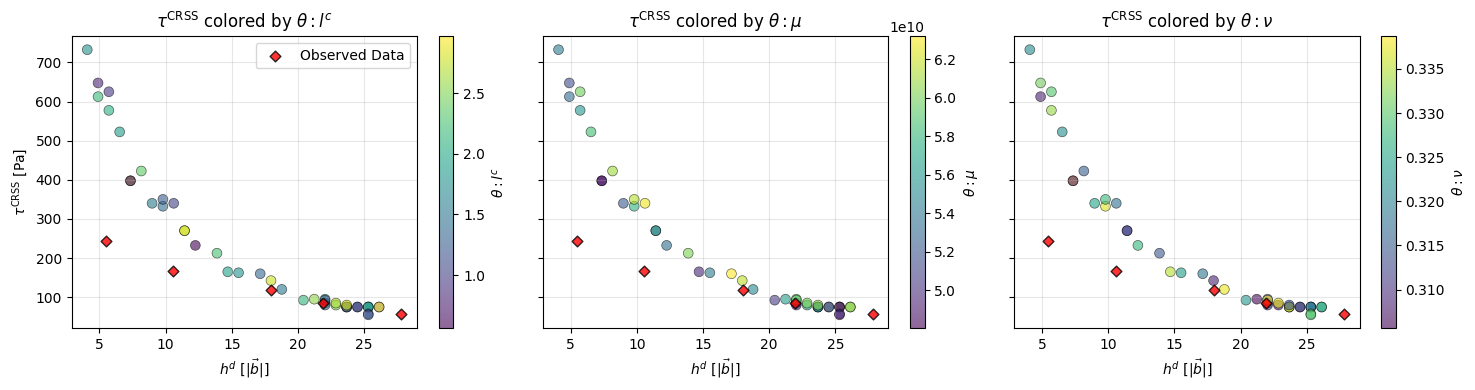}
    \caption{Plot of $\tau^{CRSS}$ in both MD (Observed Data) and DDD computed for a simple dipole microstructure. Colorbars indicate the sampled input parameters for each simulation.  The x-axis indicates the vertical separation distance between the glide planes of each segment of the dipole.}
    \label{fig:crss_distance}
\end{figure}

Figure \ref{fig:crss_distance} demonstrates that for dipole height $h^d>18|\vec{b}|$, the DDD model predictions are in relatively close agreement with the MD observations; however, for $h^d<18|\vec{b}|$, there is a significant discrepancy in the model predictions against the observed MD trends. This is caused by a dislocation core interaction $E_{SRC}$ which MD captures, yet DDD neglects in its formulation, based on linear elasticity with infinitesimal displacements. The energy of interacting dislocations takes the form of Equation \ref{eq:dislocationInteractionEnergy}.

\begin{equation}
    E_{d-d}=E_{EL}+E_{SRC}
    \label{eq:dislocationInteractionEnergy}
\end{equation}
where $E_{d-d}$ is the interaction energy between dislocations, $E_{EL}$ is a long-range elastic interaction, and $E_{SRC}$ is a short-range core interaction energy. Each component of $E_{d-d}$ is accounted for in MD, with explicit core geometries and interactions between individual atoms. The missing core component in DDD explains the systemic increase in the model discrepancy relative to observation data in the low $h^d$ regime. Physically, we can attribute the discrepancy between MD observations and DDD simulations to a deficiency of the elastic interaction kernel used to compute dislocation-dislocation interactions. 

Such model form discrepancies naturally lend themselves to interpretation via systemic analysis using Bayesian inference and the additive discrepancy provided in the KOH formalism. The $\delta_\eta$ GP as shown in Equation \ref{eq:SoAcal} would learn a functional form to describe inadequacy of the elastic interaction kernel which could serve as a correction factor if carefully applied to the DDD simulator; however, posterior distributions for the $\theta_i$ parameters would never converge to acceptable margins. For convergence to occur, the span of simulation data $y_{sim}(x_i)$ needs to envelope $y_{obs}(x_i)$, and a singular value for $\theta_i$ independent of $x_i$ would be exposed by the posterior. 

The proposed methodology has no singular $\theta_i$, as the $\theta_i$ posterior is now updated across the application domain. The $\theta_i$ posterior is generated in this case by  reinterpreting the physical meaning of the discrepancy $\delta_\eta\rightarrow\delta_\theta$. If one accepts that the model physics are sufficient to describe the observation data, and that the calibration parameters are expected to drift as a function of the application domain, the model-form error can be addressed through manipulation of the input parameters without the need to correct the model physics. The MD-DDD calibration is a natural application of such a development because one can interpret the high density of defects or close proximity of dislocations as a local variation in the material parameters. Experimental studies of a magnesium alloy demonstrate how stress-strain behavior of samples changes with dislocation density \cite{lee_dynamic_2018}. While such phenomena are usually explained by Taylor hardening laws \cite{noauthor_taylors_nodate} and Kocks-Mecking dislocation multiplication models \cite{keller_kocks-mecking_2017}, the phenomenological nature of these models does not address whether or not local changes in elastic properties are present or if the short range core interactions between dislocations can be considered as local changes to material properties. 

% however, improvements to the KOH formalism can be made to address this specific issue. Accounting for the discrepancy as a function of the application domain, or the $\tau^{CRSS}$ as a function of the dipole separation distance, directly within the model emulator GP $\eta(\cdot,\cdot)$ is a novel improvement to the KOH formalism. The described problem, regarding the calibration of $\tau^{CRSS}$ is a natural application for evaluating such methodology improvements. 

\subsection{Convergence of DDD simulations}

The selected DDD simulator, MoDELib \cite{po_phenomenological_2016}, utilizes a gauss-quadrature interpolation to determine important physical quantities along the dislocation line. As with any discrete-point continuum field solver, convergence of the simulation results ought to be evaluated prior to generating viable datasets. In predicting the critical stress necessary to drive dislocation motion, the most critical value for which convergence must be checked is the number of periodic images utilized to approximate a fully periodic simulation cell. The implemented periodicity utilizes a Ewald summation formalism as a truncation parameter, meaning that the user must prescribe an appropriate number of periodic images to achieve a convergent result. Testing the convergence of $\tau^{CRSS}_{DDD}$ for the given simulation cell shows that $N_{PBC}\geq 8$ to attain a converged result. 
% Figure \ref{} demonstrates how $\tau^{CRSS}_{DDD}$ converges as a function of $N_{PBC}$ 

Directly accounting for eight images spread radially about the simulation cell results in $16^3$ considered simulation cells, which becomes computationally intractable for larger simulations with multiple dislocations. For the relatively simplistic microstructure shown in the continuum frame of Figure \ref{fig:MD-DDDschematic}, computing dynamics utilizing $N_{PBC}=8$ is feasible, but still computationally expensive. A benefit of approaching the calibration problem utilizing GP regression is that the $\eta(\cdot,\cdot)$ GP can act as a model emulator, which can eliminate the need to continually run DDD in order to predict $\tau^{CRSS}$. 
% More work needs to be done to ensure that periodic approximations are not computationally limiting; however, such developments are considered to be outside the scope of the current work. 

\subsection{Gathering MD Observations}
\label{subsec:MDobservations}

MD simulations of the critical stress serve as the ground-truth observation data in this study due to its low cost (relative to physical experiments) and physical fidelity. The utilized simulation cell is derived from earlier work \cite{nahavandian_rate_2025}. The simulation cells are oriented such that $\vec{x}=[011]$, $\vec{y}=[1\bar{1}1]$, $\vec{z}=[21\bar{1}]$. Edge dislocations are generated into the simulation cell using Atomsk \cite{Atomsk} aligned with $\vec{z}$ and Burgers vector in $\vec{x}$, such that the total dislocation length is $153.177 \mathring{A}$. The dimensions in $\vec{x}$ and $\vec{y}$ are $154.820 \mathring{A}$ and $318.07 \mathring{A}$ respectively. The interatomic potential for Cu developed by Mishin \cite{mishin_structural_2001} is used to compute atomic interactions. 

To quantify the critical stress necessary to drive dislocations over one another, MD simulations are conducted whereby a force is applied to a layer of two atomic planes at the top and bottom of the cell in the direction of the Burgers vector resulting in an applied shear stress $\tau _{xy}$. The position of the dislocation cores is tracked throughout the simulation trajectory. Upon reaching a critical stress value, the dislocations begin to glide continuously and the global structure becomes unstable. The value of the resolved stress on the glide plane at the onset of simulation cell instability defines the observed $\tau^{CRSS}$. 
% \textcolor{red}{ENRIQUE, LETS DISCUSS THIS}

\subsection{Latin hypercube sampling of DDD simulations}
\label{subsec:DDDsampling} 
% In the DDD simulator, the energy barrier associated with $\tau^{CRSS}$ can be approximated utilizing a pseudo-NEB algorithm which computes the interaction energies between dislocations at discrete intervals along a fixed trajectory. The critical stress can be back-computed from these pseudo-NEB calculations, or 
A simplistic method for calculating $\tau^{CRSS}$ in DDD is to compute directly via the bisection algorithm, whereby multiple DDD runs are instantiated at varying stress levels in accordance with finding a minimum value for $\tau^{CRSS}$. Provided a prior distribution for the possible $\tau^{CRSS}$ values, the average value is selected first, and a simulation is instantiated where the resolved stress on the slip plane, $\tau^{RSS}=\frac{\tau_{max}+\tau_{min}}{2}$. The DDD simulator then tracks the plastic distortion throughout the simulation, which is indicative of the relative motion of the dislocation dipoles. Should $\tau^{RSS}<\tau^{CRSS}$, the stress is increased by $\frac{1}{2}$ the span of the possible remaining values. Should $\tau^{RSS}\geq\tau^{CRSS}$, the stress is reduced by the same magnitude until a minimum value for $\tau^{CRSS}$ is determined with a provided resolution. The bisection search was chosen for generating data on critical stresses in DDD, due to improved interpretability of the simulation results, despite the existence of alternative numerical methods with higher convergence rates. The ability to run multiple simulations in parallel using the Palmetto Cluster \cite{antao2024modernizing} allows for fast data generation; however, the computational price is still significant enough to warrant GP surrogate treatment. 
% The effects of modifying input parameters $\{l^C,\mu,\nu\}$ is predicted to manifest more clearly in dynamic simulations of dislocation dipoles as opposed to static calculations. 

The applied stress to the DDD simulation cell was specified such that all of the stress is resolved on the glide plane in the direction of the total burgers vector. The partial dislocations being studied sum to edge character, so a stress resolved in the direction of dislocation glide will act on each partial dislocation equally. 

To facilitate the calibration procedure, large datasets of DDD data were needed to compare with MD simulation results, and to ensure future applicability of the approach abstraction is applied. These DDD datasets ought to sample the identified DDD input parameters along the entire application domain, with specific feature engineering for $\theta_i$ depending on the relevant observed physical mechanisms $y_{obs}(x)$. Significant input parameters to be evaluated when calibrating $\tau^{CRSS}_{DDD}$ are $\vec{\theta}:=\{\mu,\nu,l^{C}\}$, where $\{\mu,\nu\}$ are elastic constants utilized for solving the elastic constitutive equations (shear modulus and Poison's ratio, respectively), and $\{l^{C}\}$ is a spreading parameter utilized to avoid computational singularities. The determination of $\tau^{CRSS}_{DDD}$ is entirely independent of the effective drag of the dislocation $B^{eff}$, so no mobility parameters ought to be considered.  

To ensure uniform sampling of the application domain and chosen calibration parameters, a latin-hypercube sample was generated for each $\{x_i,\theta_i\}$ for the calibration of $\tau^{CRSS}_{DDD}$. The Scipy.stats library \cite{scipy} was utilized to facilitate the latin hypercube sampling of the parameter space, which was subsequently passed into MoDELib \cite{po_phenomenological_2016} DDD. 43 individual predictions of $\tau^{CRSS}$ were generated for the dataset shown in Figure \ref{fig:crss_distance}, which serves as the simulation data for the proposed calibration routine. 

% A visualization of the latin hypercube sampled over 43 DDD results for $\tau^{CRSS}$ is provided in the radar plot of Figure \ref{fig:radar_hypercube}

% \begin{figure}
%     \centering
%     \includegraphics[width=0.85\linewidth]{Figures/model_radar_y_crss_DDD.png}
%     \caption{A radar plot of the sampled calibration parameters used for predicting $\tau^{CRSS}$ in the DDD model. The color inside each convex hull indicates the predicted $\tau^{CRSS}(h^d)$ for each $\theta$ permutation}
%     \label{fig:radar_hypercube}
% \end{figure}

\subsection{Integrated delta formalism}
\label{subsec:intDelta}

In this section, we outline the formulation of the "integrated delta" calibration technique, whereby the inherent model-observation discrepancy is accounted for in the emulator, as shown in Equation \ref{eq:intDeltaCal}. We define modifications to the interpretation of $\delta(x)$, and how we can utilize the summation of a constant $\theta$ with the $\delta(x)$ GP as a synthetic hyperparameter for training the model emulator $\eta(x,\theta+\delta(x))$. If one considers $\theta^*$ to be defined as

\begin{equation}
    \theta^*=\theta+\delta(x)
    \label{eq:thetaStar}
\end{equation}
then the emulator model can be  macroscopically represented in the modified form of Equation \ref{eq:SoAcal} present in step \ref{step:modification} of Algorithm  \ref{alg:embedded_discrepancy}.

The embedded discrepancy formulation considers model inadequacy to be caused entirely by the omitted dependency of the calibration parameters on $x_i$. Discrepancy in this sense no longer signifies model-observation discrepancy, but rather interprets discrepancy as a structured distortion of the simulator inputs. Therefore, a  set of $\delta_\theta$ GPs are constructed to predict the parameter drift necessary to attain model-observation agreement with coupled uncertainty quantification. 

% For the inputs $(x_i,\ i=1\ldots n)$, where $(\theta\in\mathbb{R}^d)$, we denote $(\eta(x,\phi))$ as a scalar-valued function that represents the output of the emulator, where $(\phi\in\mathbb{R}^d)$. The model-observation discrepancy, $(\delta(x)\in\mathbb{R}^d)$, is considered a vector-valued GP. To account for observation errors, we assume $e_i\sim\mathcal{N}(0,\sigma^2)$. 

Algorithm \ref{alg:embedded_discrepancy} outlines the proposed procedure, which has been programmed in Python using both the Scipy  \cite{scipy} and Scikit-learn libraries \cite{scikit-learn}. The embedded discrepancy hyperparameters $\ell_{\delta_k}$ and $\sigma_{\delta_k}$ denote the correlation length scale and discrepancy variance, respectively. Each of these parameters are systematically updated using a Metropolis-Hastings \cite{robert_metropolis-hastings_2016} accept/reject step. The final parameter that is sampled is the variance in observation noise, denoted as $\sigma^2$, which represents the $\varepsilon$ term in Equation \ref{eq:intDeltaCal} in that $\varepsilon\sim\mathcal{N}(0,\sigma^2)$. This noise is sampled using a conventional Gibbs approach \cite{yildirim_bayesian_nodate}.

\begin{algorithm}[htbp]
\caption{Bayesian Calibration with Embedded Parameter Discrepancy}
\label{alg:embedded_discrepancy}
\begin{algorithmic}[1]

\Require
Observed data $\{(x_i,y_i)\}_{i=1}^{N_o}$,
simulator evaluations $\{(x_j,\theta_j,y_j^{\text{sim}})\}_{j=1}^{N_s}$,
initial parameter estimate $\theta^{(0)}$.

\State Train a Gaussian process emulator for the simulator:
\[
\eta(x,\theta)\approx \mathcal{GP}(m_\eta(x,\theta),K_\eta((x,\theta),(x',\theta')))
\]

\State Initialize latent discrepancy fields:
\[
\delta_k^{(0)}(x_i)=0, 
\qquad k=1,\dots,d_\theta
\]

\State Specify discrepancy priors:
\[
\delta_k(x)\sim\mathcal{GP}(0,K_{\delta_k}), 
\qquad 
K_{\delta_k}(x,x')=\rho_{\delta_k}
\exp\!\left(-\frac{\|x-x'\|^2}{\lambda_{\delta_k}^2}\right)
\]

\State Specify observation model:
\[
y_i=\eta(x_i,\theta^\star(x_i))+\varepsilon_i,
\qquad 
\varepsilon_i\sim\mathcal{N}(0,\sigma^2)
\]
\label{step:modification}

\State Define embedded parameter correction:
\[
\theta^\star(x)=\theta+\delta(x)
\]

\For{$t=1,\dots,T$ (MCMC iterations)}

    \For{$k=1,\dots,d_\theta$}
    
        \State Propose an update $\delta_k'(x)$ using a Metropolis--Hastings step:
        \[
        \delta_k'(x)\sim q(\delta_k'(x)\mid \delta_k^{(t-1)}(x))
        \]
        
        \State Accept or reject based on posterior ratio:
        \[
        \alpha=\min\left(1,
        \frac{p(y\mid \theta+\delta')p(\delta')}
             {p(y\mid \theta+\delta)p(\delta)}
        \right)
        \]
        
    \EndFor

    \State Update discrepancy hyperparameters 
    $(\lambda_{\delta_k},\rho_{\delta_k})$ via Metropolis steps.

    \State Update noise variance $\sigma^2$ via Gibbs sampling:
    \[
    \sigma^2\sim p(\sigma^2\mid y,\theta,\delta)
    \]

    \State Store posterior samples of $\delta^{(t)}$, $\sigma^{2(t)}$.

\EndFor

\State Posterior predictive distribution at input $x$:
\[
p(y(x)\mid y)=\int \eta(x,\theta+\delta(x))\,p(\theta,\delta\mid y)\,d\theta\,d\delta
\]

\end{algorithmic}
\end{algorithm}

% \subsection{MCMC sampling of embedded discrepancy GP}
% \label{subsec:mcmc}

%%%%%%%%%%%%%%%%%%%%%%%%%%%%%%%%%%%%%%%%%%
\section{Results}
\label{sec:results}

To benchmark the performance of the integrated delta formalism, we compare the posterior distributions and the predictive capabilities of the emulator against the KOH formalism presented by Gattiker \cite{gattiker_gaussian_2015}. 
%Section \ref{subsec:integratedDeltaResults} presents the results of the integrated delta formalism, while Section \ref{subsec:GPMSAresults} shows the results using a conventional approach.  
% We then utilize the physics problem whose motivation is discussed in Section \ref{subsec:motivation} to apply the integrated data formalism. The results of those calibrations appear in Sections \ref{subsec:CRSScalibration}-\ref{subsec:GammaCalibration}.

\subsection{Integrated delta calibration of MD-DDD calculations for $\tau^{CRSS}$}
\label{subsec:integratedDeltaResults}

Applying the methodology described in Section \ref{subsec:intDelta} to the dataset shown in Figure \ref{fig:crss_distance} containing $N_{sim}=43$ and $N_{obs}=5$ produces a model emulator that converges to the observation data without needing an additive correction by a discrepancy model $\delta_\eta$. Instead, the predictions made using this model emulator utilize $\theta^*$ given in Equation \ref{eq:thetaStar} as the calibration parameters sampled to make predictions. 
%Figure \ref{fig:delta_intDelta} demonstrates how $\theta^*$ drifts across the application domain. 

\begin{figure}
    \centering
    \includegraphics[width=0.65\linewidth]{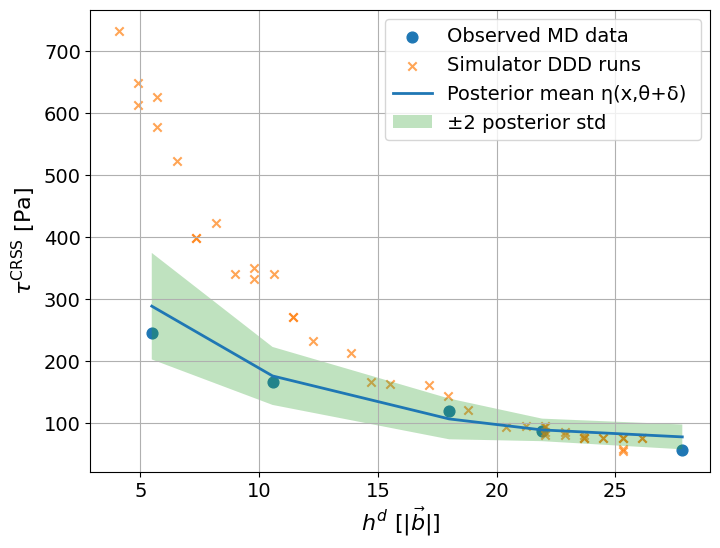}
    \caption{Predictions of the $\eta(x,\theta^*)$ emulator with coupled uncertainty.}
    \label{fig:eta_intDelta}
\end{figure}

\begin{figure}
    \centering
    \includegraphics[width=0.85\linewidth]{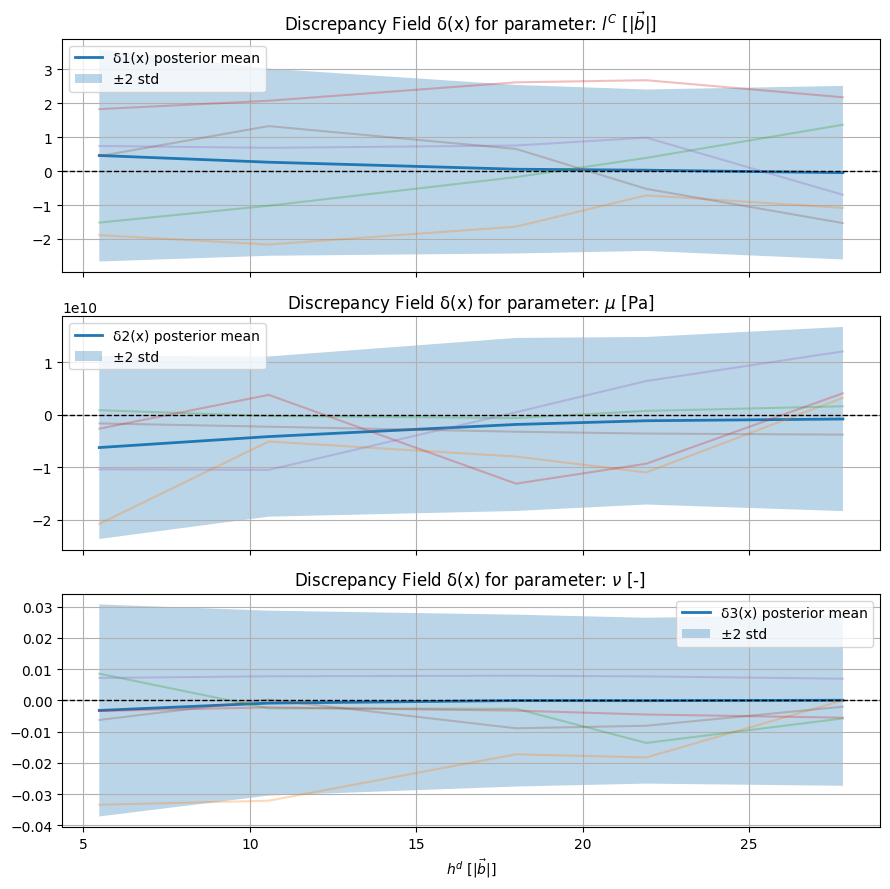}
    \caption{Parameter drift as a function of the normalized application domain. The opaque line and shaded region denote the posterior mean and standard deviation of each discrepancy field, while the translucent lines denote sampled $\delta_\theta(x_i)$ functions.}
    \label{fig:delta_intDelta}
\end{figure}

Figure \ref{fig:eta_intDelta} demonstrates the predictions of the integrated $\delta$ emulator which makes predictions based on the calibration parameters in Figure \ref{fig:delta_intDelta}. The posterior mean of the $\eta$ GP converges to MD observations, despite being trained exclusively on DDD calculations of $\tau^{CRSS}$. The trend is realized due to the extrapolated input parameters $\theta^*$. 
The results presented in Figure \ref{fig:delta_intDelta} demonstrate how the modeled material parameters ought to be corrected (drift) to ensure that the conventional elastic interaction kernel present within continuum DDD is in agreement with the discrete atomistic results of MD calculations. The short-range core effects in Cu appear to significantly decrease the effective elastic constants $\{\mu,\nu\}$ as the local density of dislocations increases (i.e. dipole separation is reduced). The core spreading parameter $\{l^C\}$, however, appears to have almost no parameter drift throughout the application domain. The non-singular theory of dislocations \cite{cai_non-singular_2006} utilizes this parameter to resolve singularities in stress-strain fields about dislocation cores. It is expected that in the limit of $h^d\rightarrow0$, that $l^C$ would have some significant change in magnitude.
%, and demonstrates how the parameter influences the stress field immediately about dislocation cores. 
One potential reason for the diminished importance of this parameter may be the prior distribution utilized in the calibration procedure. The bounds of the $l^C$ prior distribution are $[0.56,2.88]$, which span only an order of magnitude. Such bounds were chosen as 'typical' values; however, the non-physicality of the parameter lends a significant amount of uncertainty to its magnitude. Although exploring broader priors for this specific parameter may elucidate more of its influence in predicting $\tau^{CRSS}$, its lack of physical interpretability relative to elastic constants diminishes the value of such a study. 

Figure \ref{fig:delta_intDelta} also elucidates how $E_{SRC}$ can be represented in the DDD space, as a correction to the local elastic constants. As evidenced by the simulation/observation results in Figure \ref{fig:eta_intDelta}, $\lim_{h^d\to5\vec{b}-}\tau^{CRSS}_{DDD}\approx 3*\lim_{h^d\to5\vec{b}-}\tau^{CRSS}_{MD}$, which is a direct consequence of a missing $E_{SRC}$ term that would otherwise require a direct correction in the DDD elastic interaction kernel. If one considers $E_{SRC}=E_{SRC}(C_{ijkl}(\rho^{d}))$, where $C_{ijkl}$ is the $4^{th}$ order elastic tensor and $\rho^d$ is the dislocation density, and extracts the Lam\'e  parameters of $C_{ijkl}$ as a function of $\rho^d$ which is analogous in our case to $h^d$, then one can directly map the correction to the elastic constants $\{\mu,\nu\}$ to an $E_{SRC}$ term which depends on $\rho^d$ or the local density of dislocation defects. 

\subsection{GPMSA calibration of MD-DDD calculations for $\tau^{CRSS}$ }
\label{subsec:GPMSAresults}

Utilizing the KOH formalism via the GPMSA \cite{gattiker_gaussian_2015} framework on the same dataset provides insight on the model-observation error that is not available when using the integrated delta framework. The decoupling of $\delta_\eta$ from $\eta$ explicitly demonstrates which regions of the application domain contain the largest discrepancy. Using this method, the generated posterior distributions determine a singular value for each component of the $\theta_i$ parameters ; however, the results lack any conclusive information regarding optimal $\theta$ input. Despite the returned values having the maximum likelihood estimates, the parameter variation across the application domain does not result in well-converged posteriors. This is because the simulation data does not overlap observations in the low $h^d$ regime, and there likely exists no singular value for $\theta_i$ which produces a converged model with a reasonable discrepancy window. Furthermore, the $\theta_i$ parameters that may lead to tighter residuals in the low $h^d$ regime would not converge to observation data in the moderate-high $h^d$ regime, as evidenced by Figure \ref{fig:delta_intDelta}. 

\begin{figure}
    \centering
    \includegraphics[width=0.95\linewidth]{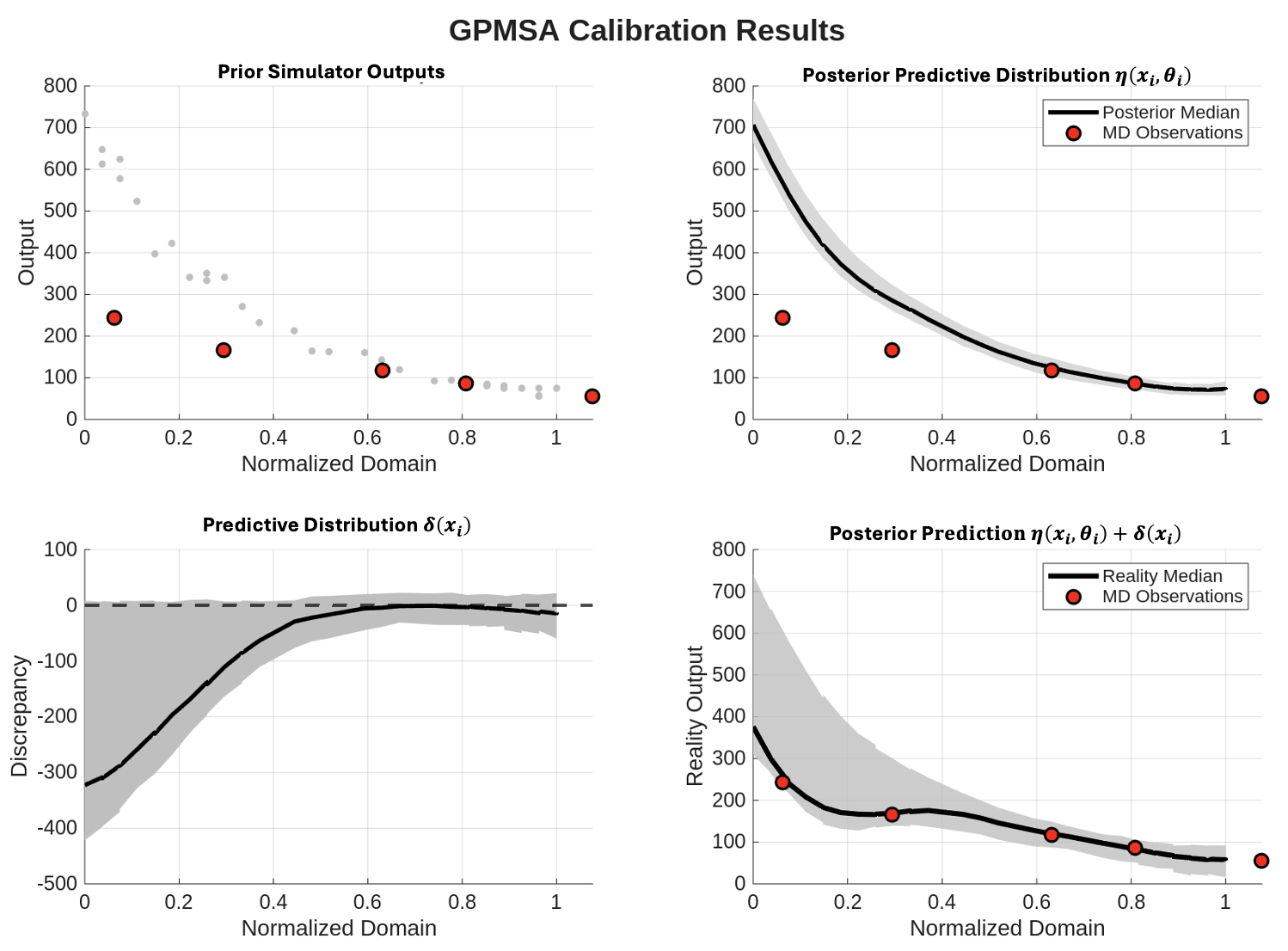}
    \caption{Model emulator and learned discrepancy from GPMSA \cite{gattiker_gaussian_2015}}
    \label{fig:gpmsa_quadchart}
\end{figure}

Despite this deficiency, the GPMSA formalism is still capable of producing a combined $\eta(x,\theta)+\delta_{\eta}(x)$ emulator converging to the observed data. Figure \ref{fig:gpmsa_quadchart} demonstrates how the summation of $\eta(x,\theta)$ (Fig. \ref{fig:gpmsa_quadchart} top-right) and $\delta(x)$ (Fig. \ref{fig:gpmsa_quadchart} bottom-left) converges with observed data (Fig. \ref{fig:gpmsa_quadchart} bottom-right).

The trained discrepancy model $\delta(x)$ captures the transition from high discrepancy in the low $h^d$ regime to the low model-observation error regime beyond $h^d\approx17\vec{b}$. In this case, the discrepancy can be entirely considered as a model inadequacy, or the lack of an explicit $E_{SRC}$ considered in DDD elastic interaction kernels. 

It should be noted that the form of Figure \ref{fig:gpmsa_quadchart}'s posterior reality prediction is most likely an overfitting stemming from noise in the given data, and not physically motivated. Specifically regarding the increasing trend at $[0.3,0.4]\in x_i^{norm}$, there is no physical motivation for increasing $\tau^{CRSS}$ as the distance between isolated, interacting dislocations increases. Analytically, the stress field emanating from each set of partial edge dislocations will decay $\sim\frac{1}{r}$, where $r$ is the radial distance away from the dislocation segment. While the application domain $h^d$ differs from $r$, in that $h^d$ is the vertical separation between the active glide planes, analytically there should be no increase of $\tau^{CRSS}$ as $h^d$ increases. The overfitting is likely driven by the large predictive distribution for $\delta_\eta$ in this region and the preceding regions. The integrated delta approach does not show the same signs of overfitting, as the smooth transition of the calibration parameters produce a monotonically decreasing trend for $\tau^{CRSS}.$

%More input on the philosophical interpretation of modeled discrepancy and confounding of uncertainty sources is provided in Section \ref{sec:discussion}.

%%%%%%%%%%%%%%%%%%%%%%%%%%%%%%%%%%%%%%%%%%
\section{Discussion}
\label{sec:discussion}

The proposed integrated delta method provides context on the source of model-form errors. However, it is only appropriate for models in which the underlying mechanisms are adequately solved via its respective numerical method. Prior information on the quality of a model is difficult to evaluate systematically without expert insight as to the various inputs of the model and their significance. In the context of the given study, the physical scale of the simulations increases the degree of uncertainty further. One cannot easily make conclusive remarks about physical phenomena at the nanometer scale without having dedicated significant time to learning the intricacies of material defects and how they influence deformation. Furthermore, treating MD simulations as concrete observations, while pragmatic, does not guarantee agreement with physical observations. The quality of MD predictions scales with respect to the quality of the utilized interatomic potential. While the authors do not anticipate much observation error for a relatively simple material such as the utilized Cu single-crystal, material complexity will add another layer of uncertainty to MD predictions that will require careful treatment. The following subsections outline potential modifications to the integrated delta formalism with cost-benefit analysis (\ref{subsec:intDeltaImprove}), and how the current results may be utilized to improve the DDD model (\ref{susbsec:DDDimprove}). 

\subsection{Proposed improvements to the integrated delta method}
\label{subsec:intDeltaImprove}

One major pitfall of the integrated delta formalism, as implemented in this study, is the inability to attribute model form error to any source outside of calibration parameter uncertainty. Equation \ref{eq:intDeltaCal} inherently assumes that all model/observation discrepancy can be described by the drift of calibration parameters across the application domain. This sacrifice may be acceptable in some cases. However, when simulation frameworks do not adequately capture the physical trends of observations, an explicit consideration of discrepancy bias as an additive corrective term ($\delta_\eta$) becomes necessary. This treatment would take the form of Equation \ref{eq:modifiedIntDelta}

\begin{equation}
    y(x)=\eta(x,\theta^*)+\delta_\eta(x)+\varepsilon
    \label{eq:modifiedIntDelta}
\end{equation}
where $\theta^*=\theta+\delta_\theta(x)$, and $\delta_\eta$ and $\delta_\theta$ describe model form discrepancy and calibration parameter discrepancy, respectively. The added complexity of such a modification is that differentiating $\delta$ into two components introduces both interparameter and intraparameter confounding. The current method only deals with intraparameter confounding, which may cause issues in the event that calibration parameters are significantly correlated. The current formalism utilizes nested or deep \cite{damianou_deep_2013} GPs where $N_\delta=N_\theta$. Each $\delta_\theta$ GP is aware of the contributions of each other $\delta_\theta$ GP, hence the intraparameter confounding. To address the issues of interparameter confounding, the authors propose a multi-step calibration process which is left for future work. 

\subsection{Implementing improvements to the DDD model}
\label{susbsec:DDDimprove}

With regards to improving the current DDD simulator, the results of each calibration routine provide a number of methods for correcting the predictions of future simulations. Calculating $\tau^{CRSS}$ is a relatively simplistic use of the DDD model, as its mesoscale representation of material defects naturally lends its uses to the computation of higher-order mechanisms such as macroscale flow rules \cite{akhondzadeh_dislocation_2020}. One would be remiss, however, to fit such rules without first ensuring model agreement with either theory or observation data for metrics like $\tau^{CRSS}$. 

The simplest way to correct the DDD simulator based on the calibration results would be to modify the effective elastic constants as a function of radial dislocation density (or proximity). The necessary modifications could be extracted from Figure \ref{fig:delta_intDelta}. One drawback of such an approach is that the modifications to the elastic constants are likely dependent on both the material and dislocation character, necessitating a much larger DDD dataset to fully account for all the possible dislocation characters (edge/screw/mixed). Even with character considered in the application domain, the results are still material dependent, and the process would need to be repeated to fully calibrate the dislocation behavior for a given crystal. The different core structures of dislocations in different materials may bear some similarities dependent on the crystal structure of the material system but that does not guarantee that the elastic parameters will drift in the same way as a function of radial defect density.

Furthermore, the primary set of equations being solved within the DDD simulator are Green's functions, which are used to compute the local stress-strain at discrete points in the simulation cell. One of the fundamental assumptions of using the Green's function is that the elastic constants do not vary in the spatiotemporal domain, which raises concern with the proposed correction method. If one interprets $E_{SRC}=E_{SRC}(\rho^d,C_{ijkl})\approx E_{SRC}(h^d,l^C,\mu,\nu)$, then one could consider an 'effective' set of elastic constants $\{\mu,\nu\}$ that modify the evaluated Green's function to agree with atomistic data. These effective elastic constants are an approximation to a necessary modification to the conventional Green's function formalism, which may account for spatially varied elastic constants. 

A more complex method for improving the DDD model would be to utilize the results of the KOH $\delta_{\eta}$ discrepancy model and apply a correction term to the elastic interaction kernel directly. Such a technique would lack the interpretability of creating effective elastic coefficients, but nevertheless account for the missing $E_{SRC}$ in DDD simulations. The method would likely suffer the same pitfalls as the implementation of the integrated delta results, in that more verification would be required to verify dislocation character and crystal structure/material dependence. The proposed implementation of calibration results to improve the DDD model is left as future work. 
%Another avenue of future work pertaining to the calibration of the DDD model will be to follow the methodology and utilize the results of \cite{nahavandian_rate_2025} in calculating and calibrating the rate of thermal activation for defect-obstacle bypass.

%%%%%%%%%%%%%%%%%%%%%%%%%%%%%%%%%%%%%%%%%%
\section{Conclusions}

In this work, we have presented a novel approach to Bayesian model calibration, with the goal of refining mesoscale material defect models to atomistic observations. The modification to the KOH formalism was necessitated by significant model form error in the predictions of a critical stress between DDD and MD that could be explained by calibration parameter drift. The model form error stems from physical interactions that are present in the MD computations but absent from the DDD predictions of dislocation trajectories. Through accounting for the discrepancy $\delta$ between the MD and DDD simulator directly within the $\eta$ emulator, the trend of $\tau^{CRSS}_{MD}(h^d)$ was recovered without the need for an external discrepancy GP. Future work aims to improve the DDD model utilizing integrated discrepancy, modifications to the integrated discrepancy, and with the overarching goal of propagating information to materials models at even larger length and time scales. 

\section*{Acknowledgments}

The authors acknowledge support from the CU-SRNL collaboration which funded this research venture.

The authors acknowledge the intellectual contributions of Dr. Giacomo Po and Dr. Nikhil Admal, and appreciate their ongoing collaboration.

This research used in part resources on the Palmetto Cluster at Clemson University under National Science Foundation awards MRI 1228312, II NEW 1405767, MRI 1725573, and MRI 2018069. The views expressed in this article do not necessarily represent the views of NSF or the United States government.
%%%%%%%%%%%%%%%%%%%%%%%%%%%%%%%%%%%%%%%%%%
% \section{Patents}

% This section is not mandatory, but may be added if there are patents resulting from the work reported in this manuscript.

%%%%%%%%%%%%%%%%%%%%%%%%%%%%%%%%%%%%%%%%%%
% \vspace{6pt} 
% \section*{References}
\newpage
\bibliographystyle{unsrtnat}
\bibliography{Box_refs}

$\,$

$\,$

% \begin{thebibliography}{99}

% \bibitem{1} Spiegel, M. R. (1981). Theory and problems of Advanced Calculus: Si (metric) edition. McGraw-Hill. 

% \end{thebibliography}
\end{document}